\documentstyle[twocolumn,floats,pre,aps,graphicx]{revtex}

\begin{document}

\draft

\title{Clustering and fluidization in a one-dimensional granular system:
molecular dynamics and direct-simulation Monte Carlo method}

\author{Jos{\'e}  Miguel Pasini\thanks{Electronic address: jpasinik@cec.uchile.cl}
and Patricio Cordero\thanks{URL: http://www.cec.uchile.cl/cinetica/} }
\address{Departamento de F{\'\i}sica, Facultad de Ciencias F{\'\i}sicas y
Matem\'aticas, Universidad de Chile, Santiago, Chile}

\maketitle

\begin{abstract}
We study a 1--D granular gas of point-like particles not subject
to gravity between two walls at temperatures $T_{\text{left}}$ and
$T_{\text{right}}$. The system
exhibits two distinct regimes, depending on the normalized temperature
difference $\Delta = (T_{\text{right}} - T_{\text{left}})/(T_{\text{right}} +
T_{\text{left}})$: one completely
fluidized and one in which a cluster coexists with the fluidized gas.
When $\Delta$ is above a certain threshold, cluster formation is fully
inhibited, obtaining a completely fluidized state. The mechanism that
produces these two phases is explained. In the fluidized state the
velocity
distribution function exhibits peculiar non-Gaussian features. For
this state, comparison between integration of the Boltzmann equation
using the direct-simulation Monte Carlo method and results stemming from
microscopic Newtonian molecular dynamics gives good coincidence,
establishing that the non-Gaussian features observed do not arise from
the onset of correlations.
\end{abstract}


\section{Introduction}

Granular systems have been extensively studied due both to the
theoretical challenges they present (for a recent review see~\cite{kadanoff99})
and to the applications of industrial importance that spring from the rich
phenomena
they exhibit (see~\cite{jaeger92,jaeger96} and references therein).
These systems are characterized by an energy loss in collisions.
This loss is at the base of many interesting phenomena, such
as inelastic collapse\cite{mcnamara92,mcnamara94}, where the particles
collide infinitely often in finite time, and
clustering (for a sample of theoretical, simulational, and experimental
approaches see~\cite{goldhirsch93,kudrolli97b,falcon99,petzschmann99}).
Different methods for keeping the system from collapsing have been devised,
such as subjecting the particles to Brownian
forces\cite{puglisi98,puglisi99}, and forcing through the boundaries by
putting the system in a box with one or more thermal-like walls (see for 
example~\cite{du95,grossman96b,grossman97,zhou98,zhou98b,ramirez99a,ramirez00a}).
This work focuses on the latter method.

Being one of the simplest types of forcing, several
authors\cite{du95,grossman96b,zhou98,zhou98b,ramirez99a,biben99}
have studied a one-dimensional system in a box with one or two heated 
(stochastic) walls. Of these, \cite{du95,grossman96b,zhou98,zhou98b}
study cluster formation, although~\cite{zhou98,zhou98b} are not strictly
one-dimensional.

This article studies a quasielastic one-dimensional 
system not subject to gravity between two thermalizing walls. We focus on
two control parameters: the total inelasticity parameter $qN\equiv N(1-r)/2$,
where $N$~is the number of particles and $r$~is the restitution coefficient,
and the externally imposed temperature gradient. The parameter~$qN$ has been shown to be
relevant for the quasielastic
system\cite{grossman96b,ramirez99a,biben99,mcnamara93,bernu94}.
By varying these
parameters we determine the region in parameter space where clustering is
fully inhibited, obtaining a fluidized state. We present a singular feature
of the distribution function for the clustering regime, and then study how
this feature is modified for the fluidized state.

In~\cite{du95}
the authors study a one-dimensional system of point-like particles between
an elastic and a heated wall. They emphasize that a cluster inevitably forms
away from the heated wall, regardless of how elastic the system is (as long
as it is not perfectly elastic). They also study the same system, but with
both walls expelling the particles with a fixed velocity. In this case they
find that the cluster forms away from the walls and roams slowly about
the system, with two groups of fast particles connecting the cluster with
the ``heated'' walls.

In~\cite{grossman96b} the same system is studied for different types of
boundary conditions at the heated wall. The stochastic boundary condition
studied has the form of a power of the velocity times the ``thermal''
condition (the one that produces a Maxwell-Boltzmann distribution in the
elastic case). The authors show that when the power that multiplies the
thermal condition is positive the test-particle equation (derived from
the Boltzmann equation) has a steady-state solution. Thus the thermal case
does not have a steady state and develops a cluster away from the heated
wall. The mechanism for the growth of the cluster is explained and verified
numerically.

In~\cite{zhou98} a similar system is studied: a long thin
pipe of inelastic hard disks with heated walls (at the same temperature)
at the ends of the pipe and
periodic side walls. The pipe is thin enough for the particle order to be
preserved. The probability distribution for the distance between the central
particles is studied. This distribution gives a markedly denser system near
the center than in the elastic case, although the limit to the elastic case
is smooth, unlike the strictly one-dimensional case
of~\cite{du95,grossman96b}. In~\cite{zhou98b} the same author studies the
velocity correlations that this system develops as inelasticity is increased,
showing that a consistent description must take these correlations into
account.

In this paper we revisit the one-dimensional system of $N$~point-like
particles interacting via collisions that conserve momentum but
dissipate kinetic energy. To fix notation, the particle velocities after a
collision are given by
\begin{equation}
c_1' = q c_1 + (1 - q) c_2, \;\;\; c_2' = (1 - q) c_1 + q c_2 ,
\end{equation}
where $c_i$ is the velocity of particle~$i$ before a collision,
and $q = (1-r)/2$, being $r$ the restitution coefficient. For the
elastic case ($r=1$) the particles simply exchange velocities. Since
the particles are point-like, the system is then indistinguishable from
a system in which the particles do not interact.

This one-dimensional system is interesting because dissipation
is the first order correction to a free gas. Besides,
results for the one-dimensional system have been found to have
unexpected relevance for higher-dimensional  problems. For example, in
two-dimensions the particles involved in inelastic collapse lie 
roughly on a line\cite{mcnamara94}. Also, the
dissipation-induced temperature gradients calculated in~\cite{ramirez99a} for
the one-dimensional case
inspired the authors to look for dissipation-induced Rayleigh-B\'enard-like
convection for a two-dimensional system without an externally imposed
temperature gradient\cite{ramirez00a}.


For a system with one thermal wall and open on the other side, under the
influence of gravity, the quasielastic system may be kept
fluidized\cite{ramirez99a,biben99}: any cluster that starts to form is forced
against the thermal wall, where it evaporates. In~\cite{ramirez99a} the
test-particle equation\cite{du95,grossman96b,mcnamara93}---which is the
1--D Boltzmann equation where the limit $N \rightarrow \infty$ is
taken, but keeping $qN$ fixed---is successfully
applied, with close matching of theory and simulations
even at the level of the distribution function.

\begin{figure}[tb]
\centering
\includegraphics[width=0.45\textwidth,clip]{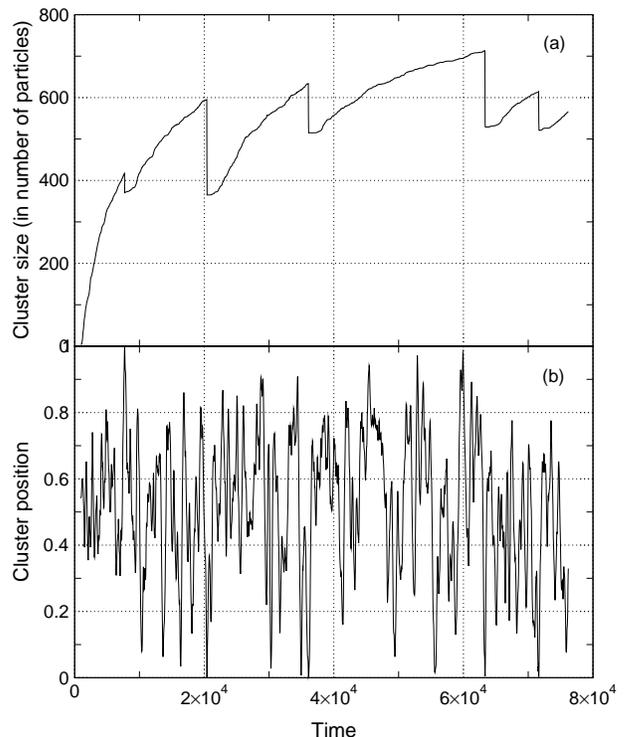}
\caption{
Subfigure~(a) shows the time evolution of the number of particles inside the
cluster for a system of $N=1000$ particles between two walls at the same
temperature for $qN=0.01$. Note the plateau after each fall.
Subfigure~(b) shows the trajectory of the cluster for the same time
interval. The walls are placed at $x=0$ and $1$. The temperature at the
walls is unity, and thus a unit of time measures how long it takes for a
particle with the thermal speed to cross the system.}
\label{fg:cluster_T0T0_qN001_N1000} 
\end{figure}

\begin{figure}[tb]
\centering
\includegraphics[width=0.45\textwidth,clip]{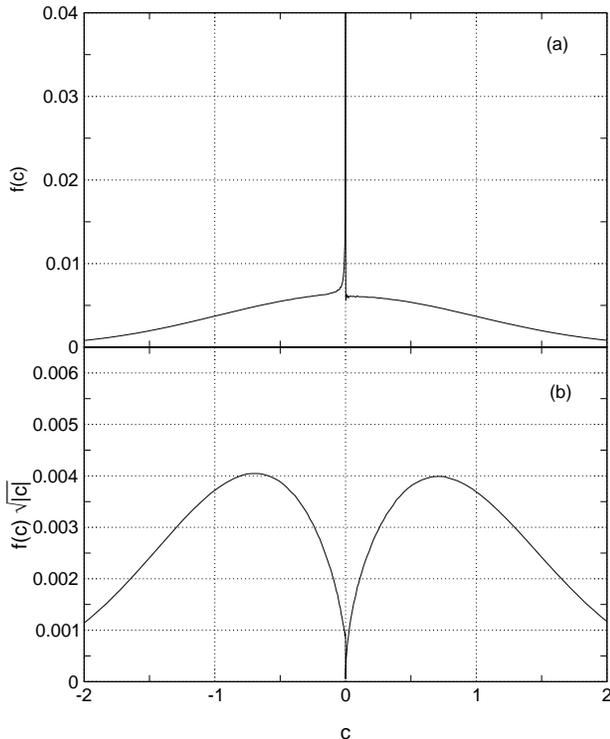}
\caption{Subfigure~(a) shows the velocity distribution function at the left wall
for the system referred to in Fig.~\ref{fg:cluster_T0T0_qN001_N1000}. The 
distribution for particles reaching the wall ($c < 0$) exhibits a sharp peak.
Subfigure~(b) shows the same distribution multiplied by~$\sqrt{|c|}$ to show
that the peak behaves like $|c|^{-1/2}$.
The microscopic velocity~$c$ is measured in units of the thermal speed.}
\label{fg:T0T0_distrib_pared_qN001}
\end{figure}

The one-dimensional system under study is left to evolve between
two thermal walls at temperatures $T_{\text{left}}$ and~$T_{\text{right}}$,
with $T_{\text{left}} \leq T_{\text{right}}$. We
define the parameter
\begin{equation}
\Delta \equiv \frac{T_{\text{right}} - T_{\text{left}}}{T_{\text{right}} +
T_{\text{left}}}
\end{equation}
to quantify how far from the symmetrical case is the system. Under the
same conditions, an elastic system has a perfectly bimodal velocity
distribution with global homogeneous temperature equal to
$\sqrt{T_{\text{left}} T_{\text{right}}}$. For the sake of comparison,
we simulate systems with $\sqrt{T_{\text{left}} T_{\text{right}}} = 1$.
Here $\Delta = 0$ represents a symmetrical
setting, while $\Delta = 1$ represents an infinitely strong
temperature gradient.


As in~\cite{du95}, for $\Delta = 0$ a cluster unavoidably forms away from the thermal
walls. After forming, the cluster performs an apparently random walk about the
center, growing in size (and therefore mass) while the rest of the
system grows more rarefied. With the decrease in density of the
surrounding gas and the increased inertia of the cluster, an eventual
collision with one of the walls is to be expected. When this happens part
of the cluster evaporates, and what is left of it is expelled from the
wall (see Fig.~\ref{fg:cluster_T0T0_qN001_N1000}), thus restarting the
growth process. Thus not only is the system highly clumped, but also in
a non-steady state. Nevertheless, the gas that is far from the
random-walk zone has a well-defined time average for the distribution
function, as is seen in Fig.~\ref{fg:T0T0_distrib_pared_qN001}. A
noteworthy feature of this distribution is that it exhibits apparently
singular behavior for slow velocities.

Setting $\Delta \neq 0$ the symmetry of the system is broken.
For $\Delta  \ll 1$ the cluster performs a slightly asymmetric
random walk, spending more time near the colder wall, and therefore
colliding more often with it. Thus the cluster cannot grow as much as
it did in the symmetric case before colliding with a wall. By
increasing $\Delta$ the cluster-wall collision
frequency grows, even obtaining short ``windows'' in which the cluster
completely evaporates. By further increasing~$\Delta$ 
these windows grow larger until a point is reached where no
cluster forms. In this fashion a totally fluidized state is
achieved, which may be tractable with the dissipative Boltzmann
equation. However, the distribution function obtained from the molecular
dynamics (MD)
simulations exhibits a peculiar non-Gaussian feature for slow
velocities. This feature is a smoothed version of the apparent
singularity of the symmetric case. To discern whether this feature is
due to correlations or is present before they settle in, we
compared Newtonian molecular dynamics results with those obtained through
direct-simulation Monte Carlo (DSMC), which neglects correlations. The results
agree very well, except when the system approaches the clustering
regime.

\section{Simulation method}

We simulate the system through event-driven molecular
dynamics\cite{marin93} and through direct-simulation Monte
Carlo\cite{bird94}.  The direct-simulation Monte Carlo procedures use
the null-collision technique\cite{koura86} where, overestimating the
collision frequency (using the maximum relative velocity within a
cell), the number of collisions to be {\em attempted} is calculated
through a Poisson process.  In the next step the collisions are
attempted, choosing at random two particles within the cell, and making
them collide with a probability proportional to their relative velocity.
Most of the molecular dynamics and DSMC simulations were done with
$N=1000$ particles.

In the MD simulations we detect clusters using a geometric criterion:
we consider chains of particles that are nearer than a critical distance
(in our case~$10^{-6}$ and~$10^{-5}$, to be certain that the conclusions
are independent of the choice). The system length is one, and
with a thousand particles the mean distance between neighbors for a
homogeneous system is~$10^{-3}$. Thus we detect particles that are
uncommonly near by three orders of magnitude. We discard chains of length
three or less, since they may be random encounters. Measuring the {\em total}
length of the cluster we have found that on average it is of the order
of~$10^{-5}$; thus the choice of $10^{-5}$ as link-link distance is much
larger than the true distance between them.

As already stated, the boundary conditions are such that the (homogeneous)
temperature of the corresponding elastic system (equal to
$\sqrt{T_{\text{left}} T_{\text{right}}}$) is one.

\section{Results}

\begin{figure}[tb]
\centering
\includegraphics[width=0.45\textwidth,clip]{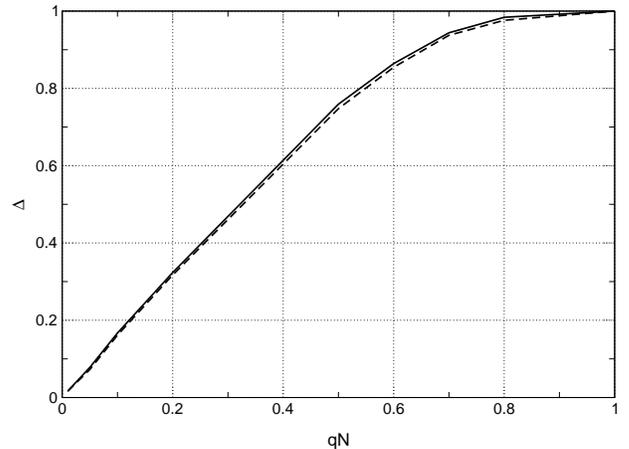}
\caption{Limiting values of $\Delta \equiv (T_{\text{right}} - T_{\text{left}})/
(T_{\text{right}} + T_{\text{left}})$ as a function of $qN$ for $N=1000$.
$\sqrt{T_{\text{left}} T_{\text{right}}} = 1$ throughout.
The solid curve shows the lowest possible value~$\Delta$ can take without
detecting clusters. The dashed curve shows the largest possible value~$\Delta$
can take with easy cluster detection. The small region between the curves
represents a zone where clusters appear erratically.}
\label{fg:clustering_threshold}
\end{figure}

\begin{figure}[tb]
\centering
\includegraphics[width=0.45\textwidth,clip]{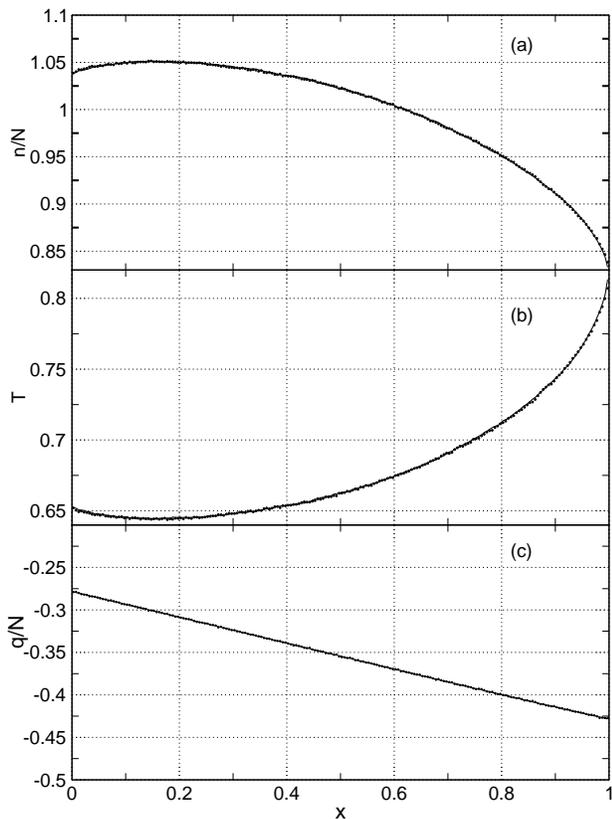}
\caption{Number density, temperature, and heat flux profiles for $N=1000$,
$qN=0.1$, and $T_{\text{left}} = 0.5$ ($\Delta=0.6$). The solid line represents results from
a molecular dynamics simulation, while the dots represent results from 
a Monte Carlo simulation. The density and temperature are related by
$p = nT$ and, since momentum is conserved and the system is stationary,
the pressure is constant throughout the system. The density and temperature
profiles are almost symmetrical because the normalized density is very close
to unity.}
\label{fg:qN01Ti05_n_T} 
\end{figure}

\begin{figure}[tb]
\centering
\includegraphics[width=0.45\textwidth,clip]{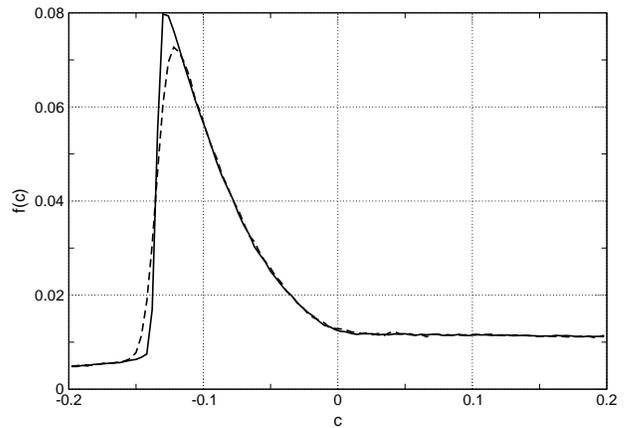}
\caption{Velocity distribution function at the cold boundary (left wall)
for the system referred to in Fig.~\ref{fg:qN01Ti05_n_T}. The solid line
represents results from a molecular dynamics simulation, while the dashed
line represents results from a Monte Carlo simulation.}
\label{fg:qN01Ti05_q_dist}
\end{figure}

\subsection{Clustering regime}

Figure~\ref{fg:cluster_T0T0_qN001_N1000} shows the non-steady state of a
granular system between two walls at the same temperature for $qN=0.01$.
A cluster forms away from the walls, performing a random walk of varying
amplitude. When the cluster reaches a wall, part of it evaporates, and the
growth process begins anew.

As is usual for the quasielastic case, we relabel the particles when they
collide. This enables us to visualize this system as a group of barely
interacting particles passing through each other.

The picture for cluster evolution, as explained before, is the following:
the cluster grows because the slowest particles, due to the asymmetry of
the distribution function, drift towards the cluster\cite{grossman96b}.
As it grows, the density of the gas surrounding it
decreases, with the consequent saturation in growth. Thus we have a
``Brownian particle'' of increasing mass moving in an increasingly
rarefied medium. This ``particle'' will be increasingly less affected by
the surrounding medium, until it can no longer be kept away from the
walls.

The cluster moves several orders of magnitude slower than the
thermal speed (three orders of magnitude in
Fig.~\ref{fg:cluster_T0T0_qN001_N1000}). Upon reaching a wall, the 
front liners strike the wall and are expelled by it much faster than
the other cluster members. These particles pass through the cluster,
transferring momentum to it, as described in~\cite{du95}. Thus these
fast particles push the cluster away from the wall, where it can absorb
particles again. The fast particles, however, no longer belong to the
cluster.

Since the slowest particles in the gas are the ones that will be absorbed
by the cluster\cite{grossman96b}, it is the number of slow particles in the
gas that will determine the cluster's growth rate. After the cluster strikes a
wall, the expelled particles will be fast particles, and they will not
contribute to the growth of the cluster during the time it takes for
the gas to cool down again: the only particles available for absorption
are the ones that were available before the cluster-wall collision.
This explains why the cluster keeps growing at approximately the same
rate it did before the collision. After the gas has cooled down, the
growth rate returns to its normal value. This is the end of the plateau
seen in Fig.~\ref{fg:cluster_T0T0_qN001_N1000} after each cluster-wall
collision.

To quantify the evaporation process we proceed as
in~\cite{kadanoff99,du95}:
as soon as the first particle belonging to the cluster reaches a wall,
it is expelled with a speed much higher than the cluster velocity; thus we may
consider
the ideal situation of a cluster of $M$ particles at rest being stricken
by a fast particle with velocity~$v$ (in this case $v \approx 1$). 
After colliding with the
first particle in the cluster, the new fast particle's speed will be $(1-q)$.
Thus, after traversing the cluster, the fast particle's velocity will be
$(1-q)^N$. Since momentum is conserved in collisions, the center of mass of
the cluster will have acquired a speed of
\begin{equation}
v_{\text{CM}} = \frac{1 - (1-q)^N}{N} = \frac{1 -
\left(1-\frac{qN}{N}\right)^N}{N}  .
\end{equation}
Considering the case $N \gg 1$ with fixed~$qN$, as in~\cite{ramirez99a}, we
may simplify this expression to
\begin{equation}
v_{\text{CM}} \approx \frac{1 - e^{-qN}}{N}  .
\end{equation}
By further considering the case $qN \ll 1$ we get $v_{\text{CM}}
\approx q$. In this limit, if the cluster reaches the wall with
velocity~$v_0$ and is expelled from it with velocity $v_1$, the number $m$ of
particles evaporated will satisfy $|v_1 - v_0| \approx m q$.
For the situation shown in Fig.~\ref{fg:cluster_T0T0_qN001_N1000}, $v_0$
and~$v_1$ are
typically of the order of~$10^{-3}$ and $q = 10^{-5}$, hence the number of particles evaporated
will be of the order of~$10^2$.

Even if the system is in a non-steady state, the gas at the walls (far
from the random-walk zone) has a well-defined time average for the
distribution function.
The distribution function at the left wall is shown in
Fig.~\ref{fg:T0T0_distrib_pared_qN001}. There is an apparent
singularity for slow velocities. The distribution is asymmetric as it
should, since the particles leaving the wall ($c>0$) follow a
Gaussian distribution. Figure~\ref{fg:T0T0_distrib_pared_qN001}
also shows the distribution multiplied by $\sqrt{|c|}$. Since the limit
of $\sqrt{|c|}\,f$ for $c\rightarrow 0^-$ is finite and nonzero, we
conclude that the distribution function exhibits a singularity that
behaves like $|c|^{-1/2}$ for slow velocities.

As shown in~\cite{grossman96b}, when $\Delta = 0$ (the symmetric case) the
distribution shown in Fig.~\ref{fg:T0T0_distrib_pared_qN001} is not
a solution of the steady-state test-particle equation:
\begin{equation}
c \, \partial_x f(x,c,t) = qN \, \partial_c \left[f(x,c,t) M(x,c,t)
\right]  , \label{eq:steadystateTPE}
\end{equation}
where
\begin{equation}
M(x,c,t) = \int_{-\infty}^\infty f(x,c',t) (c - c') |c - c'| dc'  .
\end{equation}
To establish this, let us study the behavior for small~$c$ of a solution of
this equation. Assume that, for small~$c$, $f(x,c\approx 0) \approx f_0(x)
c^{-\alpha}$, with $\alpha < 1$ in order to have a finite density in the
vicinity of~$x$. Furthermore, assume that $M(x,c\approx 0) \approx M_0(x)
c^\beta$. Inserting this behavior in Eq.~(\ref{eq:steadystateTPE}) we obtain
\begin{equation}
\partial_x f_0 \approx qN M_0 f_0 (\beta - \alpha) c^{\beta - 2}.
\end{equation}
Thus, in order to keep $f_0$ (the amplitude of the singularity) finite,
we must have either $\beta = \alpha$ or $\beta \geq 2$. Integrating the
distribution of Fig.~\ref{fg:T0T0_distrib_pared_qN001} we obtain $\beta = 1$.
Since $\beta < 2$, we must have $\alpha = 1$. But this corresponds to a
nonintegrable distribution, and therefore the distribution cannot
be steady.

\subsection{Inhibition of cluster formation}

Figure~\ref{fg:clustering_threshold} shows the regions in $(\Delta,qN)$-space
where clustering is inhibited for $N=1000$. As is to be expected, as the
inelasticity increases, a stronger temperature gradient is necessary to
inhibit cluster formation.

To discern whether the non-Gaussian features of the velocity distribution
function are derived from correlations in the system we compared results from
MD simulations (full Newtonian dynamics) with results from DSMC simulations
(no velocity correlations assumed). Figures \ref{fg:qN01Ti05_n_T}
and~\ref{fg:qN01Ti05_q_dist} show this comparison for a case far from the
clustering threshold ($\Delta = 0.6$ and $qN=0.1$). The temperature of the
left and right walls are chosen so that the global temperature for the elastic
case (equal to $\sqrt{T_{\text{left}} T_{\text{right}}}$)
is one. The curves match almost exactly.

At the level of the distribution function, the results also match
closely. The peculiar non-Gaussian feature of the distribution function
is clearly seen in Fig.~\ref{fg:qN01Ti05_q_dist}. There is some slight
mismatch near the peak.

\begin{figure}[tb]
\centering
\includegraphics[width=0.45\textwidth,clip]{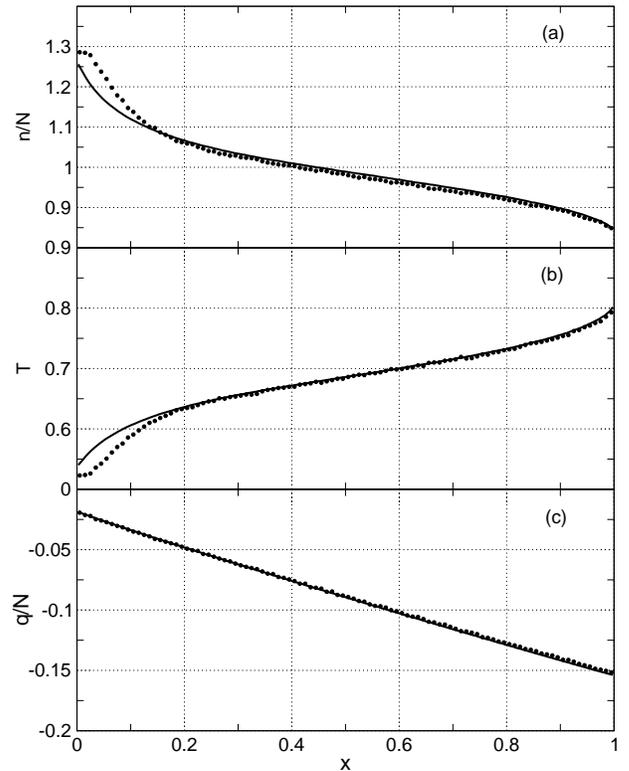}
\caption{Number density, temperature, and heat flux profiles for $N=1000$,
$qN = 0.1$, and $T_{\text{left}} = 0.84$ ($\Delta=0.173$). The solid line 
represents results from a molecular dynamics simulation, while the dots
represent results from a Monte Carlo simulation.} \label{fg:qN01Ti084_n_T}
\end{figure}

For the fluidized case, since momentum is conserved and the system is
stationary, the pressure is constant throughout the system. The
number density and the granular temperature calculated here are related
by $p = nT$ ($T$ in energy units). Thus when the normalized
density~$n/N$ varies little throughout the system ($n/N \approx 1 +
\epsilon(x)$), the normalized temperature is
\begin{equation}
\frac{T}{T_0} = \frac{p/n}{p/n_0} = \frac{n_0/N}{n/N} = \frac{1}{1+\epsilon(x)}
\approx 1 - \epsilon(x),
\end{equation}
thus obtaining the nearly symmetric profiles seen in Figs.~\ref{fg:qN01Ti05_n_T}
and~\ref{fg:qN01Ti084_n_T}.

\begin{figure}[tb]
\centering
\includegraphics[width=0.45\textwidth,clip]{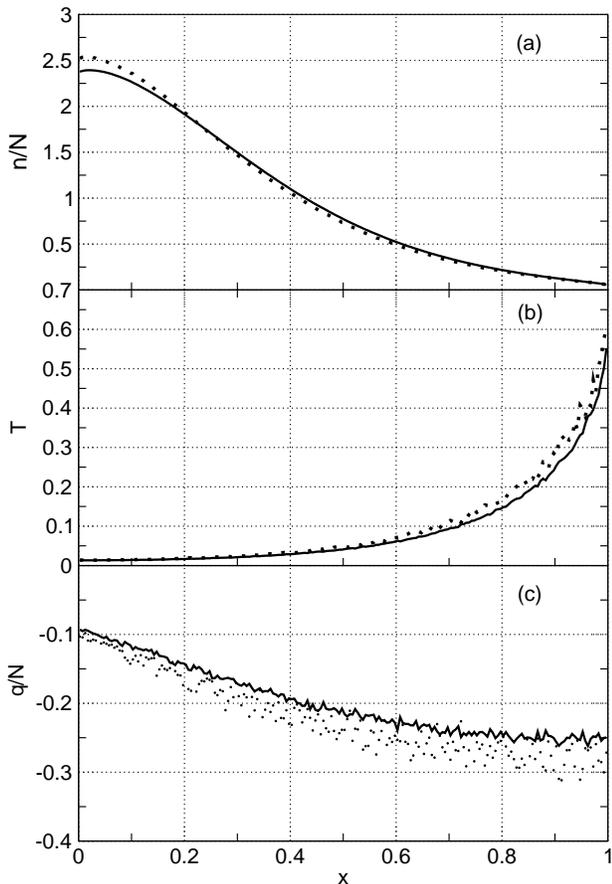}
\caption{Number density, temperature, and heat flux profiles for $N=1000$,
$qN = 0.5$, and $T_{\text{left}} = 0.01$ ($\Delta = 0.9998$). The solid line
represents results from a molecular dynamics simulation, while the dots
represent results from a Monte Carlo simulation.}
\label{fg:qN05Ti001_n_T}
\end{figure}

\begin{figure}[tb]
\centering
\includegraphics[width=0.45\textwidth,clip]{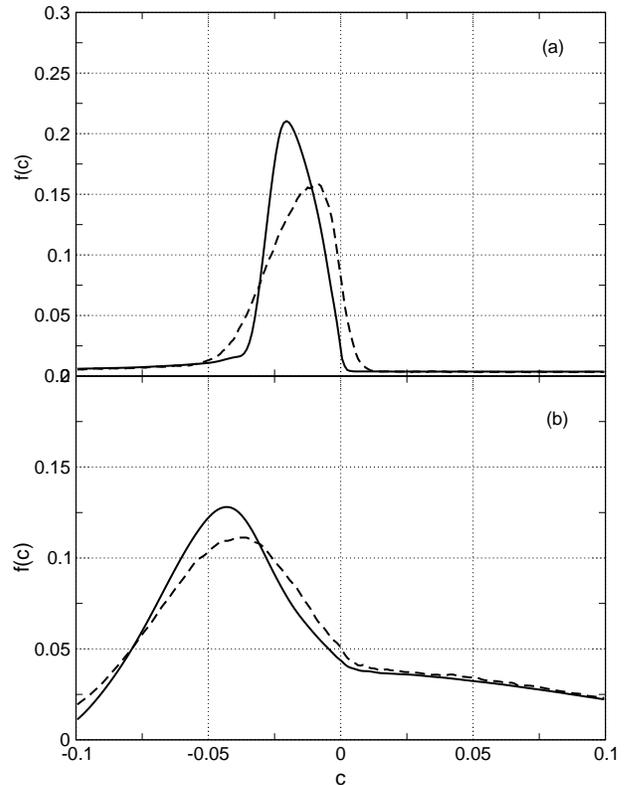}
\caption{Velocity distribution functions at the cold boundary (left wall).
Subfigure~(a) corresponds to the system referred to in
Fig.~\ref{fg:qN01Ti084_n_T}, while subfigure~(b) corresponds to the system of
Fig.~\ref{fg:qN05Ti001_n_T}.
The solid line represents results from a molecular dynamics (MD) simulation,
while the dashed line represents results from a Monte Carlo (DSMC) simulation.
The MD results have been rescaled so that the area under the
MD and DSMC curves is the same. There is a systematic overpopulation of slow
particles in the distributions obtained from Monte Carlo simulations.}
\label{fg:qN05Ti001_q_dist}
\end{figure}

Figures \ref{fg:qN01Ti084_n_T}, \ref{fg:qN05Ti001_n_T},
and~\ref{fg:qN05Ti001_q_dist} compare the MD and DSMC results for cases
near cluster formation. The non-Gaussian feature of
the distribution function shows a systematic deviation for DSMC
simulations: there is overpopulation for slow velocities. This is
explained by considering that the DSMC method, like the
Boltzmann equation, neglects correlations. When the system approaches
the clustering regime, increased dissipation induces correlations which
tend to make the particles collide less\cite{zhou98b}. In DSMC
these correlations are neglected, with the corresponding systematic
overestimation in the collision frequency. This
overestimation results in a lower temperature of the system about the
density peak.

\section{Conclusions}

We have shown that a system not subject to gravity between thermal walls
unavoidably reaches a non-steady state when the walls are at the same
temperature.  A cluster forms in the bulk, slowly roaming about the system
while absorbing particles.  As it grows, the amplitude of the random walk
increases, until at last the surrounding gas cannot keep the cluster away
from the walls. When the cluster reaches a wall, a part of it is ejected
by the wall {\em through the cluster} (relabeling the particles on
collisions), effectively pushing the cluster
away from the wall, and leaving it to grow again.

Most of the time the cluster is far from the walls.  Thus measuring the
distribution function at a wall is measuring the distribution function
of the gas that surrounds the cluster.  This distribution function has
a well-defined time average, and exhibits apparently singular behavior for
slow particles, diverging like~$|c|^{-1/2}$.

Imposing an external temperature gradient forces the cluster against the
colder wall, inhibiting its growth. Increasing the temperature difference
leads to a system in which the cluster never forms: the system is completely
fluidized. The distribution function of the gas exhibits peculiar
non-Gaussian features: a smooth  version of the aforementioned singularity.
Therefore, any attempt at solving the Boltzmann equation through moment
methods must consider this feature in the initial {\em ansatz}, as is done
for the problem of an infinitely strong shock wave in~\cite{grad69}
and~\cite{cercignani99}. In fact, a solution for this problem was
attempted using the four moment method of~\cite{liu61}. As mentioned
in~\cite{clause88}, the fourth balance equation could not be freely
chosen when the boundary conditions were symmetric: some choices gave
undefined results. As is natural, by not including the non-Gaussian
feature in the {\em ansatz} for this calculation we obtained absurd
results, such as higher temperature in the middle of the
system than near the walls.

We compared molecular dynamics with direct-simulation Monte Carlo. Agreement
between these two methods shows that the non-Gaussian feature of the
distribution function may be predicted by the dissipative Boltzmann
equation. As the system approaches cluster formation, correlations settle
in. These correlations reduce the collision frequency among
particles. DSMC neglects these correlations, and thus overestimates the
number of collisions. This exaggerates the effects of dissipation, producing
steeper profiles.

\acknowledgments

We thank Rodrigo Soto, Aldo Frezzotti, and Rosa Ram{\'\i}rez
for helpful discussions. This work has been partially funded by
{\em Fundaci\'on Andes} through a doctoral scholarship, {\em Fondecyt}
through grants 2990108 and~1000884, and by {\em FONDAP} through
grant~11980002.

\end{document}